\documentclass[twocolumn,reprint,superscriptaddress,amsmath,amssymb,aps,nofootinbib]{revtex4-1}

\usepackage{dcolumn}
\usepackage[pdftex]{graphicx}
\usepackage{bm}
\usepackage{array}
\usepackage[utf8]{inputenc}
\usepackage{amsmath}
\usepackage{amsfonts}
\usepackage{bbold}
\usepackage{color}
\usepackage{graphicx}
\usepackage{float}
\usepackage[T1]{fontenc}
\usepackage[english]{babel}
\usepackage{amsthm}
\usepackage{makeidx}
\usepackage{amssymb}

\usepackage{braket}
\usepackage{mathtools}

\usepackage[english]{varioref}
\usepackage{url}
\usepackage[pagebackref]{hyperref}
\restylefloat{table}
\usepackage{multirow}
\usepackage{xcolor}
\usepackage{booktabs}
\begin{document}
\title{
The TopFlavor scheme in the context of $W'$ searches at LHC
}
\author{Roberta Calabrese}
\email{rcalabrese@na.infn.it}
\affiliation{Dipartimento di Fisica ``Ettore Pancini”, Universit\`a degli studi di Napoli Federico II, Complesso Univ. Monte S. Angelo, I-80126 Napoli, Italy and
INFN - Sezione di Napoli, Complesso Univ. Monte S. Angelo, I-80126 Napoli, Italy}
\author{Agostino De Iorio}
\email{agostino.deiorio@unina.it}
\affiliation{Dipartimento di Fisica ``Ettore Pancini”, Universit\`a degli studi di Napoli Federico II, Complesso Univ. Monte S. Angelo, I-80126 Napoli, Italy and
INFN - Sezione di Napoli, Complesso Univ. Monte S. Angelo, I-80126 Napoli, Italy}
\author{Damiano Fiorillo}
\email{dfgfiorillo@na.infn.it}
\affiliation{Dipartimento di Fisica ``Ettore Pancini”, Universit\`a degli studi di Napoli Federico II, Complesso Univ. Monte S. Angelo, I-80126 Napoli, Italy and
INFN - Sezione di Napoli, Complesso Univ. Monte S. Angelo, I-80126 Napoli, Italy}
\author{Alberto Orso Maria Iorio}
\email{albertoorsomaria.iorio@unina.it}
\affiliation{Dipartimento di Fisica ``Ettore Pancini”, Universit\`a degli studi di Napoli Federico II, Complesso Univ. Monte S. Angelo, I-80126 Napoli, Italy and
INFN - Sezione di Napoli, Complesso Univ. Monte S. Angelo, I-80126 Napoli, Italy}
\author{Gennaro Miele}
\email{miele@na.infn.it}
\affiliation{Dipartimento di Fisica ``Ettore Pancini”, Universit\`a degli studi di Napoli Federico II, Complesso Univ. Monte S. Angelo, I-80126 Napoli, Italy and
INFN - Sezione di Napoli, Complesso Univ. Monte S. Angelo, I-80126 Napoli, Italy}
\affiliation{Scuola Superiore Meridionale, Universit\`a degli studi di Napoli ”Federico II”, Largo San Marcellino 10, 80138 Napoli, Italy}
\author{Stefano Morisi}
\email{smorisi@na.infn.it}
\affiliation{Dipartimento di Fisica ``Ettore Pancini”, Universit\`a degli studi di Napoli Federico II, Complesso Univ. Monte S. Angelo, I-80126 Napoli, Italy and
INFN - Sezione di Napoli, Complesso Univ. Monte S. Angelo, I-80126 Napoli, Italy}

\begin{abstract}
Many extensions of the Standard Model predict the existence of new charged or neutral gauge bosons, with a wide variety of phenomenological implications depending on the model adopted. The search for such particles is extensively carried through at the  Large Hadron Collider (LHC), and it is therefore of crucial importance to have for each proposed scenario quantitative predictions that can be matched to experiments.
In this work we focus on the implications of one of these models, the TopFlavor Model, proposing a charged $\text{W}^\prime$ boson that has preferential couplings to the third generation fermions. We compare such  predictions to  the ones from the so called Sequential Standard Model (SSM), that is used as benchmark, being  one of the simplest and most commonly considered models for searches at the LHC. We identify the parameter space still open for searches at the LHC, and we show that the cross sections for the processes $pp \to \text{W}^\prime\to \tau\nu$ and $pp \to \text{W}^\prime\to tb$ in the TF assume different values with respect to the SSM as a function of the particle mass and width, and that the TF has realizations that would not be allowed in the SSM and not yet excluded by data. This study makes the case for further searches at the LHC, and shows how a complete and systematic model independent analysis of $\text{W}^\prime$ boson phenomenology at colliders is essential to provide guidance for future searches. 
\end{abstract}

\maketitle

\section{introduction}
\label{sec:intro}
The Standard Model (SM) successfully describes three fundamental interactions, and its predictions are in excellent agreement with data. It is however known that, even putting aside gravitational interactions,  the SM cannot be the ultimate fundamental theory.
Among the main clues leading to such conclusion are the observation of baryon asymmetry, the neutrino oscillation phenomena, and the dark paradigm. 
The SM also presents issues of self-consistency, like the Higgs hierarchy problem, that, while not corresponding to a specific observation, undermine its robustness as a fundamental theory valid at all energy regimes. Finally, statistically significant evidence of Nature's behavior deviating from the SM has been mounting over the recent years in several sectors of particle physics, and that could be explained in the context of new physics scenarios. In the b-physics sector, flavor anomalies have been spotted at b-factories, BaBar and Belle, and at LHCb in the D*~\cite{babarlfvd,bellelfvd,lhcblfvd} and K*~\cite{babarlfvk,bellelfvk,lhcblfvk} decay ratios, possibly pointing towards lepton flavor universality violation \cite{Hiller:2017bz}. The anomalous magnetic moment of the muon ($g-2$) at Brookhaven National Laboratory \cite{Bennett_2006} and Fermilab National Accelerator Laboratory \cite{Abi:2021gix} has also been found to deviate significantly from SM expectations.

All the above points to the fact that the SM must be extended by some new theory, that can be built with  top-down as well as bottom-up approaches.

Examples of top-down approaches are several gauge extensions of the SM that are inspired by Grand Unified Theories like the Pati-Salam model \cite{pati1994lepton}. This model is based on the gauge group 
$SU(4)_c \times SU(2)_L \times  SU(2)_R$, also referred to as ({\bf4},\,{\bf2},\,{\bf2}), which is a maximal subgroup of $SO(10)$ grand unification group. 
In a bottom-up approach, other gauge symmetries can be considered, like the left-right $SU(3)_c \times SU(2)_L \times  SU(2)_R\times U(1)$, also referred to as L-R model \cite{rabindra1986unification, Corrigan:2015kfu},
and the $SU(3)_c \times SU(3)_L \times  U(1)_Y$, also referred to as ({\bf3},\,{\bf3},\,{\bf1}) model \cite{pisano19923, buras2013anatomy,boucenna2015predicting, cabarcas2010flavor, cogollo2012novel}. Other possibilities have been proposed, inspired by the extension of the Higgs sector rather than the introduction of new symmetries, like Little Higgs models \cite{goh2007little, schmaltz2004simplest,Antipin_2013} and Twin Higgs models  \cite{ahmed2018heavy, chacko2006twin}. 
All these SM extensions have a common feature: the prediction of new $\text{W}^\prime$ and $\text{Z}^\prime$ gauge bosons, \cite{langacker2009physics} in analogy with the W and Z gauge bosons of the SM. 
In the current letter, we will focus on $\text{W}^\prime$ boson phenomenology, giving an overview of the models that can foresee its introduction, and focusing on deriving measurable predictions on a specific one, named TopFlavor (TF) Model.

The paper is organized as follows: Section~\ref{sec:wprime} serves as a brief overview on the most common $\text{W}^\prime$ models. In Section~\ref{sec:topflavor} we will introduce the general properties of the TF model studied in this work. In Section~\ref{sec:pheno} we report the phenomenological implications for collider searches with a comparison with the SSM case, and in Section~\ref{sec:results} we draw quantitative predictions for LHC searches. In section ~\ref{sec:conclusions} we report our conclusions.

\section{$\text{W}^\prime$ models overview}
\label{sec:wprime}
The couplings of $\text{W}^\prime$ with SM particles, fermions, scalars, and vectors, depend on the specific gauge model.
The new gauge boson interactions with fermions can be written in a general way as 
\begin{equation}
\begin{split}
\mathcal{L}^{\rm eff}&=\frac{V_{f_if_j}}{2\sqrt{2}}g_w \bar{f_i} \gamma_\mu [\alpha_R^{f_if_j}(1+\gamma^5) + \\
&+\alpha_L^{f_if_j}(1-\gamma^5) ]W^{\prime \mu}f_j+h.c.,
\label{eqeff}
\end{split}
\end{equation}
where $V_{f_if_j}$ is the analogue of the Cabibbo-Kobayashi-Maskawa (CKM) matrix if $f_i$ ed $f_j$ represent quarks, while for leptons  $V_{f_if_j} = \delta_{ij}$ and $g_w$ is the coupling with the $\text{W}^\prime$ \cite{collaboration2012search}.

The parameters $\alpha_R^{f_if_j}$ and  $\alpha_L^{f_if_j}$ are free in a {\it model independent} analysis, while specific $\text{W}^\prime$-models correspond to specific choices them. 

The {\it sequential standard model} (SSM) described in Ref.~\cite{altarelli1989searching} is defined to have the same couplings to fermions as the SM W boson, leading to $g_w=e/\sin \theta_W$, $\alpha_L^{f_if_j}=1$, and $\alpha_R^{f_if_j}=0$ for $i,j=1,2,3$.\footnote{Note that in this model this holds true also for the $Z^\prime$ boson, and it holds true both for the vertex with fermions and the ones involving other vector bosons and the Higgs, namely $\text{W}^\prime f\bar f$, $\text{Z}^\prime f f^\prime$, $\text{W}^{\prime\, \pm} \text{W}^\mp \text{Z}$, $\text{Z}^\prime \text{W}^+ \text{W}^-$.}
It is worth mentioning that the SSM is {\it not expected} in the context of any gauge theory unless new scalars and fermions are assumed to extend the SM beside the $\text{W}^\prime$ boson~\cite{langacker2009physics}. Indeed the inclusion of a new $\text{W}^\prime$ boson requires to extend the gauge group with, for instance, an extra $SU_{\rm new}(2)$ group. 
In order to couple to the $\text{W}^\prime$ boson, the SM fermions, both quarks and leptons, must transform under the new $SU_{\rm new}(2)$. The minimal extension of the weak gauge group by means of a new $U(1)$ provides no $\text{W}^\prime$ boson, but gives a $\text{Z}^\prime$ one. Another feature of these models is that they require new scalar fields, since it is necessary to reproduce the SM with a spontaneous symmetry breaking (SSB) of the new symmetry group.

LR gauge models provide a possible example of such an extension, based on the $SU(2)_L \times  SU(2)_R\times U(1)$ gauge group \cite{rabindra1986unification, Corrigan:2015kfu}, and give  
$\alpha_R^{f_if_j}=\alpha_R$ and  $\alpha_L^{f_if_j}=\alpha_L$  where $\alpha_{L,R}$ are arbitrary parameters.

Another possible extension is the class of models based on a ({\bf3},\,{\bf3},\,{\bf1}) gauge symmetry. In this case, the $\alpha_R^{f_if_j}$ and  $\alpha_L^{f_if_j}$ assignment depends on the details of the model. In fact, within the ({\bf3},\,{\bf3},\,{\bf1})  model there is some arbitrariness in the assignment of the matter field in order to complete the irreducible representation of $SU(3)_L$, namely the anti-triplet $\overline{\bf 3}$. However most of ({\bf3},\,{\bf3},\,{\bf1}) models provide $\alpha_L^{f_if_j}\ne 0$ and $\alpha_R^{f_if_j}=0$ in analogy with the SSM. For instance, the Lagrangian of the model presented in Ref.~\cite{pisano19923} contains 
\begin{equation}
\begin{split}
    \mathcal{L}\supset& -\frac{g}{\sqrt{2}}(
    \bar{\ell}_L^c\gamma^\mu \nu_{\ell\,L} \text{W}^{\prime\,+}_\mu+
    \bar{J}_{1\,L}\gamma^\mu u_{L} \text{W}^{\prime\,+}_\mu \\ 
    &-\bar{q}_{i\,L}\gamma^\mu J_{i\,L} \text{W}^{\prime\,+}_\mu
    +h.c.)\,,
\end{split}
    \label{331example}
\end{equation}
where $g=e \sqrt{1+3\sin^2 \theta}/\sin \theta$, $\tan^2 \theta \approx 11/6$, i=2,3, and $J_{1,2,3}$ are new quarks with exotic charges. It is important to notice that the quarks in the Lagrangian of Eq.~(\ref{331example}) are not the mass eigenstates.

In conclusion, the effective Lagrangian reported in Eq.~(\ref{eqeff}) is the most general one that parametrises the coupling of a $\text{W}^\prime$ boson with fermions. Nevertheless it is not satisfactory from a theoretical point of view in its most general form. To compute phenomenological, observable predictions, one must often reduce to a subset of parameters compatible with the conditions listed above.

In particular, for what concerns couplings to fermions, experimental searches often focus on two benchmark cases:
\begin{eqnarray}
\text{W}^\prime_L: && \quad \alpha_L^{f_if_j}=1\,,\quad \alpha_R^{f_if_j}=0\,,\nonumber\\
\text{W}^\prime_R: && \quad \alpha_L^{f_if_j}=0\,,\quad \alpha_R^{f_if_j}=1\,,\label{eqcms}
\end{eqnarray}
both of them have $g_w=e/\sin\theta_W$.
While the first case is exactly the SSM introduced in Ref.~\cite{altarelli1989searching}, the second one is its right-handed version that is a special case of the LR model. The two cases in Eq.\,(\ref{eqcms}) do not cover the full extent of possible models that could actually appear in Nature. Other combinations of parameters could be allowed, motivated by different theoretical models or assumptions, resulting in a wider parameter space to explore at the LHC or future colliders.

In this work, we will explore the phenomenological implications of a third class of $\text{W}^\prime$ models, denoted as TopFlavor Model\,\cite{li1993gauge, malkawi1996model}, whose key assumptions are significantly different with respect to the ones leading to Eq.\,(\ref{eqcms}). In particular, we will show how a vast portion of the parameter space available to this model is not yet excluded by the LHC.
We will evaluate the production rate of  $pp \to \text{W}^\prime \to \tau \nu$ and $pp \to \text{W}^\prime \to tb$ processes,  showing that it can range down to approximately a fifth of the one of the SSM. 
\vspace{5.mm}

\section{TopFlavor model}
\label{sec:topflavor}

The most general realization of TF model is based on the gauge group \cite{PhysRevLett.47.1788}:
\begin{equation}
SU(3)_C\times SU(2)_1 \times SU(2)_2 \times SU(2)_3 \times U(1)_Y\,,\label{eqg1}
\end{equation}
where the  {\it $i^{th}$} flavor generation transforms as a doublet under $SU(2)_i$ and as a singlet under the other $SU(2)_j$ with i,j in (1,2,3), and $i\ne j$. The presence of three separate gauge groups for each family leaves considerable freedom in the realization of the TF model. In this work, we consider the one given in Refs.~\cite{lee1998phenomenological, Muller_1996, cao2016interpreting, muller1997separate, PhysRevLett.47.1788,li1993gauge}, where the first two generations transform under the same $SU(2)_{12}$ and the third family, the heaviest one, transforms under $SU(2)_{3}$. Under such assumption, the gauge group reported in Eq.~(\ref{eqg1}) reduces to:
\begin{equation}
SU(3)_C\times SU(2)_{12} \times SU(2)_3 \times U(1)_Y\,.\label{eqg2}
\end{equation}
Such a group can be obtained from the more general one in Eq.~(\ref{eqg1}) by a SSB mechanism \cite{PhysRevLett.47.1788}. The model also requires to extend the scalar sector with two new fields: $\Phi$, that transforms as a doublet under $SU(2)_{12}$, and $\Sigma$, that is  a bi-doublet under  $SU(2)_{12} \times SU(2)_3 $. We can write the bi-doublet scalar fields as:
\begin{equation} 
\Sigma=\left( \begin{array}{cc} \sigma+i\pi_3&i\pi_1+\pi_2\\ i\pi_1-\pi_2&\sigma-i \pi_3 \end{array} \right), 
\end{equation}
where $\pi^i$ and $\sigma$ are real fields, and the doublet scalar field as:
\begin{equation}
\Phi=\left(\begin{array}{c} \Phi^+\\ \Phi^0\end{array}\right). 
\end{equation} 
The transformation rules of these fields are:
\begin{equation} \label{eq:transformation}
    \begin{split}
    &\Sigma \to g_{12}\,\Sigma\,g_3^\dagger, \\
    &\Phi \to g_{12}\,g_Y\,\Phi,
    \end{split}
\end{equation}
where $g_{12}\in SU(2)_{12}$, $g_3\in SU(2)_3$, and $g_Y\in U(1)_Y$.

Similarly to the SM, the degrees of freedom of all the real fields present in the scalar sector, except for $\text{Re}(\Phi^0)$ and $\sigma$, are converted into the longitudinal component of the gauge bosons.
The vacuum expectation values (v.e.v.s) of these fields are\footnote{Both expectation values $u$ and $v$ can be taken real after a suitable gauge transformation as in Eq.~\ref{eq:transformation}. For more general classes of model this is could not be possible (for example see L-R model).}
\begin{equation}
\begin{split}
\braket{\Sigma}&=\left( \begin{array}{cc} u&0\\0&u\end{array} \right), \quad
\braket{\Phi}=\left(\begin{array}{c}0\\ v\end{array}\right),
\end{split}
\end{equation}
where both $v$ and $u$ are real numbers. 
The field $\Phi$ plays the same role as the Higgs field in the SM, with the difference that it couples to the third generation only.

The pattern of the SSB from the full symmetry group of Eq.\,(\ref{eqg2}) proceeds as follows: in the first step the field $\Sigma$ acquires its v.e.v. ($u \gg v$), leading to the SSB:
\begin{equation}
SU(2)_{12} \times SU(2)_3 \xrightarrow{\Sigma}  SU(2)_L;
\end{equation}
in the second step, the field $\Phi$ acquires its v.e.v. causing the SSB:
\begin{equation}
SU(2)_{L} \times U(1)_Y \xrightarrow{\Phi}  U(1)_{EM},
\end{equation}
which is the same breaking as in the SM.

The first two generations of leptons $L^{1,2}$ transform as $(2,1)$ under $SU(2)_{12} \times SU(2)_3 $ while $L^3$ transforms as $(1,2)$. On the other hand, right handed leptons are singlets under  $SU(2)_{12} \times SU(2)_3 $. Similar assignments are given for quarks.

The model also contains seven gauge bosons, corresponding to the four SM gauge bosons and the new $\text{W}^\prime$ and $\text{Z}^\prime$ bosons.

The matter content of the model is summarized in Table~\ref{TopFlavor table}.

\begin{table}[ht]
\centering
\renewcommand\arraystretch{1.3}
\begin{tabular}{cccc}
\toprule
        & $SU(2)_{12}$ & $SU(2)_3$ & $U(1)_Y$\\
\midrule
$L^3$   &   1&  2&  $-1/2$\\
$L^{1,2}$&  2&  1&  $-1/2$\\
$Q^3$   &   1&  2&  $1/6$\\
$Q^{1,2}$&  2&  1&  $1/6$\\
$u^{1,2,3}_R$&  1&  1& $2/3$\\
$d^{1,2,3}_R$&  1&  1& $-1/3$\\
$\ell^{1,2,3}_R$&  1&  1& $-1$\\
$\Sigma$&   2&  $2$   &0\\
$\Phi$& 1&  2&  $1/2$\\
\bottomrule
\end{tabular}
\caption{Matter content of the TopFlavor model. \label{TopFlavor table}}
\end{table}

Since the strong interaction sector does not change, we will omit its description.

The complete Lagrangian of the model is given for instance in Ref.~\cite{malkawi2000light} and is summarized here as:
\begin{equation}
\mathcal{L}=\mathcal{L}_B+\mathcal{L}_F+\mathcal{L}_{Y}, 
\end{equation}
where $\mathcal{L}_F$ contains the fermion components,  $\mathcal{L}_B$ contains the boson and scalar components , while $\mathcal{L}_Y$ contains the Yukawa interaction and is not reported here since it is model dependent and does not influence the phenomenological searches of $\text{W}^\prime$ at accelerators. $\mathcal{L}_F$ and $\mathcal{L}_B$ expressions are reported in the following:
\begin{equation}
\begin{split}
&\mathcal{L}_{F}= i \bar{\Psi}\gamma^{\mu}D_{\mu}\Psi=\\
&i\bar{L}_L^j\gamma_\mu \left( \partial^\mu+i\frac{g_{12}}{2}\sigma^i W_{12}^{i\mu}-i\frac{g_0}{2}B^\mu \right) L_L^j+\\
+&i\bar{Q}_L^j\gamma_\mu \left( \partial^\mu+i\frac{g_{12}}{2}\sigma^i W_{12}^{i\mu}+i\frac{g_0}{6}B^\mu \right) Q_L^j+\\
+&i\bar{L}_L^3\gamma_\mu \left( \partial^\mu+i\frac{g_{3}}{2}\sigma^i W_{3}^{i\mu}-i\frac{g_0}{2}B^\mu \right) L_L^3+\\
+&i\bar{Q}_L^3\gamma_\mu \left( \partial^\mu+i\frac{g_{3}}{2}\sigma^i W_{3}^{i\mu}+\frac{g_0}{6}B^\mu \right) Q_L^3+\\
+&i\bar{u}_R^\alpha\gamma_\mu \left( \partial^\mu+\frac{2g_0}{3}B^\mu \right) u_R^j+\\
+&i\bar{d}_R^\alpha\gamma_\mu \left( \partial^\mu-i\frac{ g_0}{3}B^\mu \right) d_R^j+\\
+&i\bar{e}_R^\alpha\gamma_\mu \left( \partial^\mu-ig_0 B^\mu \right) e_R^j,
\end{split}
\label{Lagrangiana}
\end{equation}
\begin{equation}
\begin{split}
&\mathcal{L}_B=\frac{1}{2}D_\mu\Phi^\dagger D^\mu \Phi+\frac{1}{4}\rm{Tr}(D_\mu \Sigma^\dagger D^\mu \Sigma)+\\
&-V(\Sigma; \Phi)-\frac{1}{4}W^a_{12\mu \nu}W^{a\mu \nu}_{12}+\\
&-\frac{1}{4}W^a_{3\mu \nu}W^{a\mu \nu}_3-\frac{1}{4}B_{\mu \nu}B^{\mu \nu}\,, 
\end{split}
\end{equation}
where $i=1,2$, $\alpha=1,2,3$, and the covariant derivative is defined as $D^{\mu}=\partial^{\mu}+ i g_{12} \vec T\cdot \vec W_{12}^{\mu}+i g_3 \vec T^{\prime}\cdot \vec W_{3}^{\mu}+i g_0Y B^{\mu}$.
Here $W_3^{1,2,3}$, $W_{12}^{1,2,3}$, and $B$ are the gauge bosons fields.
As usual these can be written in terms of charged and neutral bosons $W_3^{\pm}$, $W_{12}^{\pm}$, $W_3^0$, $W_{12}^0$, and $B$. We also observe that the model has five free parameters, that are $u$, $v$, $g_{12}$, $g_3$, $g_0$.

The mass matrix of the charged bosons in the basis $W_{12}^\pm$, and $W_{3}^\pm$ is given by \cite{Muller_1996}:
\begin{equation}
\mathcal{M}_{1}=\left( \begin{array}{cc}
\frac{g_{12}^2(u^2+v^2)}{4}& -\frac{g_{12}g_3u^2}{4}\\
-\frac{g_{12}g_3u^2}{4}& \frac{g_3^2u^2}{4} \end{array} \right).
\label{eqn:matrice1}
\end{equation}
On the other hand, the neutral bosons mass matrix in the basis $B$, $W_{12}^{0}$, $W_3^0$ is given by:
\begin{equation}
\mathcal{M}_{2}=
\left( \begin{array}{ccc}
\frac{g_0^2v^2}{4}&-\frac{g_{12}g_0v^2}{4}&0\\
-\frac{g_{12}g_0v^2}{4}&\frac{g_{12}^2(v^2+u^2)}{4}&-\frac{g_{12}g_3u^2}{4}\\
0&-\frac{g_{12}g_3u^2}{4}&\frac{g_3^2u^2}{4}
\end{array} \right)\, .
\label{eqn:matrice2}
\end{equation}
As expected, the matrix in Eq.\,(\ref{eqn:matrice2}) admits one massless eigenvalue, corresponding to the photon state. The diagonalisation of the matrix in Eq.\,(\ref{eqn:matrice1}) leads to a mixing between the charged bosons: the mixing angle between $W^{\pm}_{12}$ and $W^\pm_{3}$ will be denoted by $\theta'$ in the following\footnote{There is a slight difference ($\sim v^2/u^2$) between the mixing angles $\theta^\prime$ and $\phi$, the latter being the mixing angle between $W_{12}^0$ and $W_3^0$.}. Similarly, the diagonalisation of the matrix in Eq.\,(\ref{eqn:matrice2}) leads to a mixing between the neutral bosons: the mixing angle between $B$ and $W^0_{12}$ will be denoted by $\theta$.
The requirement that the coupling between the photon and the charged leptons is equal to the electric charge leads to the following relations in the limit $u^2\gg v^2$:
\begin{equation}
\begin{split}
g_0&=\frac{e}{\cos \theta}\, , \\
g_{12}&=\frac{e}{\sin \theta \cos \theta^\prime}\, , \\
g_3&=\frac{e}{\sin \theta \sin \theta^\prime}\, , 
\end{split}
\label{eqg}
\end{equation}
where $e$ is the electric charge. The three couplings are therefore not linearly independent: we can rather use as free independent parameters the quantities $\theta$, $\theta^\prime$, $u$, and $v$.  
The eigenvalues of the charged and neutral gauge mass matrices are the physical masses of the bosons, which in the limit $u^2\gg v^2$ are:
\begin{equation}
\begin{split}
M_\text{W}^2 &\simeq \frac{v^2}{4}\frac{e^2}{\sin^2 \theta}\left(1-\sin^4 \theta^\prime \frac{v^2}{u^2}\right)\, , \\
M_\text{Z}^2 & \simeq \frac{v^2}{4}\frac{e^2}{\sin^2 \theta\cos^2 \theta}\left(1-\sin^4 \theta^\prime \frac{v^2}{u^2}\right)\, , \\
M_{\text{Z}^\prime}^2  &\simeq M_{\text{W}^\prime}^2  \simeq \\ 
&\frac{e^2v^2}{4 \sin^2 \theta}\left( \tan^2 \theta^\prime + \frac{u^2}{v^2 \sin^2 \theta^\prime \cos^2 \theta^\prime }\right)\, . 
\end{split}
\label{eqn:mass}
\end{equation}

We require that the masses of the W and Z bosons agree with the experimental values, namely $M_{\text{W}}=80.379\pm 0.012$ GeV and $M_{\text{W}}/M_{\text{Z}}=0.88147\pm0.00013$ \cite{PDG}. This requirement leads to: 
\begin{equation}
\begin{array}{c}
\sin^2 \theta\simeq \sin^2 \theta_{W} =0.23, \\
v\simeq v_{SM}\simeq 246\, \text{GeV},
\end{array}
\label{eqc1}
\end{equation}
where $\theta_W$ is the Weinberg angle, and $v_{SM}$ is the v.e.v. of the SM. These relations impose two new constraints on the four free parameters $\theta$, $\theta'$, $u$ and $v$: in the end we are left with two free parameters, namely $u$ and $\theta^\prime$. 

Since our aim is to discuss the production of $tb$ and $\tau\nu_\tau$ \textit{via} virtual $\text{W}^\prime$ boson decay, we give the interaction term between the leptons and the charged bosons as:
\begin{equation}
    \begin{split}
        &\mathcal{L}\supset\frac{1}{\sqrt{2}} [h_{I,II}\left( e_L \gamma^\mu \text{W}_\mu^-\nu_{e\,L}+\mu_L \gamma^\mu \text{W}_\mu^- \nu_{\mu\,L} \right)+\\
    &+h_{III}\tau_L\gamma^\mu \text{W}_\mu^-\nu_{\tau\,L}+h_{III}^\prime \tau_L \gamma^\mu \text{W}_\mu^{\prime\,-}\nu_{\tau\,L}+\\
    &h_{I,II}^\prime\left( e_L \gamma^\mu \text{W}_\mu^{\prime\,-}\nu_{e\,L}+\mu_L \gamma^\mu \text{W}_\mu^{\prime\,-} \nu_{\mu\,L} \right)]+\text{h.c.}\,,
    \end{split}
    \label{tabellacoupling}
\end{equation}
where $h_{I,II}=g_{12}\cos \theta^\prime$, $h_{III}=g_{3}\sin \theta^\prime$, $h_{I,II}^\prime=g_{12}\sin \theta^\prime$, and $h_{III}^\prime=g_{3}\cos \theta^\prime$. An analogous expression can be obtained for the quarks. In the limit $u^2 \gg v^2$, we have that $h_{I,II}\approx h_{III}\approx g_{SM}$ (see Eq.\,(\ref{eqg})).

\section{TopFlavor model at colliders}
\label{sec:pheno}
The search for a $\text{W}^\prime$ boson at colliders is typically done by looking at the products of its decay after its production as a real state. 
More specifically, at LHC, the $\text{W}^\prime$ boson would be produced in the process $pp\to \text{W}^\prime$ with cross section $\sigma(pp\to \text{W}^\prime)$.
It subsequently would decay into quarks ($\text{W}^\prime\to {qq}^\prime$) or leptons ($\text{W}^\prime\to \ell \nu$) with branching ratios denoted as $\text{Br}(\text{W}^\prime\to {qq}^\prime)$ and $\text{Br}(\text{W}^\prime\to \ell \nu)$ respectively. 
These channels have been studied by the ATLAS and CMS Collaborations as benchmark cases~\cite{Sirunyan:2018mpc,Sirunyan:2019vgj,Sirunyan_2019,Sirunyan:2017vkm,Atlas_wtolep,Aaboud:2017yvp,Aaboud:2018vgh,Aaboud_2019,Aaboud:2018juj, CMS:2021mux}. A first set of studies of the phenomenological implications of TF models at the LHC has been conducted considering proton-proton collisions at 7 and 8 TeV in~\cite{TFRunI_1,TFRunI_2}.

As in the case of SM W boson, the couplings of the quarks with the $\text{W}^\prime$ boson have to take into account the inequality between the flavor basis and the mass basis for the quarks. In the SM this leads to the presence of the CKM matrix. If $V_u$ and $V_d$ are the matrices connecting the mass and flavor eigenstates for the up-type and down-type quarks respectively, the CKM matrix is equal to $V_{\text{CKM}}^{SM}=V_u\,V_d^\dagger$.
The unitary matrices $V_u$ and $V_d$ are not separately observable in the SM and we have freedom to chose as basis the one where $V_u= \mathbb{1}$,  is equal to the identity matrix, while $V_d^\dagger=V_\text{CKM}$ without loss of generality. In the TF model, this is not anymore true, and $V_u$ and $V_d$ are arbitrary $3\times 3$ matrices. A complete scan of the full $V_u$ and $V_d$ parameter space is beyond the scope of the present paper.
In this work we assume that $V_u=\mathbb{1}$ , while $V_d^\dagger=V_\text{CKM}$. It can be shown that with this choice the CKM matrix for the $\text{W}^\prime$ boson is:
\begin{equation}
    V_{\text{CKM}}^\prime=G\cdot V_{\text{CKM}}^{\text{SM}}+\frac{h_{III}^\prime}{h_{I,II}^\prime}R\cdot V_{\text{CKM}}^{\text{SM}},
    \label{eqn:CKM}
\end{equation}
where:
\begin{equation}
    R=\left(\begin{array}{ccc}
         0&0&0  \\
         0&0&0\\
         0&0&1
    \end{array}\right), \,
    G=\left(\begin{array}{ccc}
         1&0&0  \\
         0&1&0\\
         0&0&0
    \end{array}\right).
\end{equation}
We note that in the limit $h_{I,II}^\prime=h_{III}^\prime$ one finds $V_{\mathrm{CKM}}^{SM}=V_{\mathrm{CKM}}^\prime$.
Instead, by assuming $V_u=V_{\text{CKM}}^{\text{SM}}$ and $V_d=\mathbb{1}$ it follows: 
\begin{equation}
    V_{\text{CKM}}^\prime= V_{\text{CKM}}^{\text{SM}}\cdot G+\frac{h_{III}^\prime}{h_{I,II}^\prime} V_{\text{CKM}}^{\text{SM}}\cdot R
    \label{eqn:CKM2}
\end{equation}
that is quite different from the one in Eq.\,(\ref{eqn:CKM}). However it is possible to show that the phenomenological implications in these two extreme cases are similar and for this reason, in the following we will use the assumption of Eq.\,(\ref{eqn:CKM}). This property does not hold for all the possible choices of $V_u$ and $V_d$.

The branching fraction of $\text{W}^\prime \to tb$ and $\text{W}^\prime \to \tau \nu_\tau$ requires the knowledge of all possible partial decay widths of the $\text{W}^\prime$ boson in this model. 
It is possible to show that  the three bosons vertexes $\text{W}^\prime \to \text{WZ}$ and $\text{W}^\prime \to \text{W} \gamma$
have a null coupling as well as the four bosons vertexes $\text{W}^\prime \to \text{WWW}$, $\text{W}^\prime \to \text{WZZ}$, $\text{W}^\prime \to \text{W} \gamma \gamma$ and $\text{W}^\prime \to \text{WZ} \gamma$.
The decay channels $\text{W}^\prime \to \text{WH}$ and $\text{W}^\prime \to \text{WHH}$ are negligible compared to the ones involving fermions. The dominant partial decay widths are therefore:
\begin{equation}
    \begin{split}
        &\Gamma\left(\text{W}^\prime\to tq\right)=\frac{h_{I,II}^{\prime\, 2}}{16\pi}|V_{tq}^\prime|^2 \frac{\beta^2}{M_{\text{W}^\prime}}\left(M_{\text{W}^\prime}^2+ \frac{m^2_t}{2}\right), \\
        &\Gamma\left(\text{W}^\prime\to qq^\prime\right)=\frac{h_{I,II}^{\prime \, 2}}{16 \pi} |V_{qq^\prime}^\prime|^2 M_{\text{W}^\prime},\\
        &\Gamma\left(\text{W}^\prime\to e \nu_e\right)=\frac{h_{I,II}^{\prime\,2}}{16\pi}\frac{M_{\text{W}^\prime}}{3},\\
        &\Gamma\left(\text{W}^\prime \to \mu \nu_\mu\right)=\Gamma\left(\text{W}^\prime\to e \nu_e\right),\\
        &\Gamma(\text{W}^\prime \to \tau \nu_\tau)=\frac{h_{III}^{\prime\,2}}{16\pi}\frac{M_{\text{W}^\prime}}{3}.
    \end{split}
\label{partial decay width}
\end{equation}
where $\beta^2=1-\frac{m_t^2}{M_{\text{W}^\prime}^2}$ and $q,\,q^\prime\ne t$ in the second line. 
The resulting total decay width of the $\text{W}^\prime$ boson is:
\begin{equation}
\begin{split}
    &\Gamma_{\text{Tot}}= \frac{h_{III}^{\prime\, 2}}{16\pi} \frac{\beta^2}{M_{\text{W}^\prime}}\left(M_{\text{W}^\prime}^2+ \frac{m^2_t}{2}\right)+\\
    &+ \frac{(h_{III}^{\prime\,2}+2h_{I,II}^{\prime\,2})}{16\pi}\frac{M_{\text{W}^\prime}}{3}+\frac{\,h_{I,II}^{\prime \, 2}}{8\pi} M_{\text{W}^\prime}\,,
    \end{split}
    \label{gammatot}
\end{equation}
where we used Eq.~(\ref{eqn:CKM}) and the properties of $V_{\text{CMK}}^{\text{SM}}$. Finally the branching fractions for the $tb$ and $\tau \nu_\tau$ decays are:
\begin{equation}
\begin{split}
    &\text{Br}(\text{W}^\prime \to tb)=\\
    &\frac{\left(\frac{h_{III}^{\prime\, 2}}{16\pi}|V_{tb}^{SM}|^2 \frac{\beta^2}{M_{\text{W}^\prime}}\left(M_{\text{W}^\prime}^2+ \frac{m^2_t}{2}\right)\right)}{\Gamma_{\text{Tot}}}\,,\\
    &\text{Br}(\text{W}^\prime \to \tau \, \nu_\tau)=\frac{h_{III}^{\prime\,2}}{16\pi}\frac{M_{\text{W}^\prime}}{3\,\Gamma_{\text{Tot}}}\,.
    \label{tau}
        \end{split}
\end{equation}

\begin{figure}[ht]
\centering
\includegraphics[width=70mm]{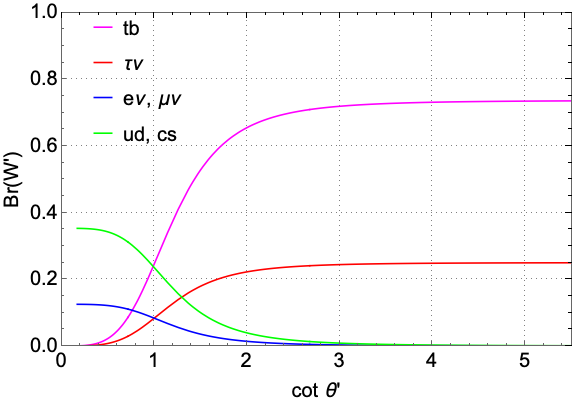}
\caption{Main branching fractions as function of $\cot \theta^\prime$}
\label{branching_plot}
\end{figure}

The main branching fractions, i.e. the ones involving decays to leptons and quarks, are shown as a function of $\cot \theta^\prime$ in Fig.~\ref{branching_plot}.
\vskip5.mm

As it is shown, for example in Ref.~\cite{ZprimeToTTBarPaper}, if the new particle was to have a non negligible decay width with respect to experimental resolutions, the sensitivity analysis to its existence could be reduced. 
In order to calculate the proton-proton production cross section, we make use of the \textsc{MadGraph5\_MC@NLO} tool in the 5 flavor scheme \cite{Alwall:2014hca}. The model~\cite{Sullivan:2002jt,Duffty:2012rf} uses the Lagrangian as reported in Eq.~(\ref{eqeff}) with the typical assumptions of the SSM: $g_w=e/\sin \theta_W$, $\alpha_L^{f_if_j}=1$, $\alpha_R^{f_if_j}=0$ and  $V_{f_if_j}=V_{\mathrm{CKM}}$. The third row and column of $V_{\mathrm{CKM}}^{SM}$ are approximated to (0,0,1). 
To simulate the Parton Density Functions (PDF) of protons, we use the NNPDF3.1\cite{Ball:2017nwa} PDF set, derived at leading order and with $\alpha_s =0.118$. 
Table~\ref{tab:wpr_xs} reports the values of the cross section and their relative uncertainty for a centre-of-mass energy of $\sqrt{s}=13$ and 14 TeV. The values in Table~\ref{tab:wpr_xs} allow us to obtain the cross sections for the TF model by multiplying them times $\tan^2\theta^\prime$ since $h_{I,II}^\prime=g_{SM}\tan \theta^\prime$.

\begin{table}[htbp]
\resizebox{!}{9.12cm}{
\begin{tabular}{llll}
\toprule
\multirow{2}{*}{Mass} & \multirow{2}{*}{Width} & \multicolumn{2}{c}{Cross section (fb)} \\
                      & & 13 TeV   & 14 TeV       \\
\midrule
\multirow{3}{*}{1000} & 10    & 109552$\pm$407   & 127176$\pm$466   \\
                      & 100   & 10939$\pm$54   & 12674$\pm$ 59    \\
                      & 200   & 5355$\pm$21    & 6177 $\pm$ 23   \\
\midrule
\multirow{3}{*}{1400} & 14    & 26036$\pm$123   & 31207$\pm$ 144   \\
                      & 140   & 2712$\pm$12   & 3246 $\pm$ 13\\
                      & 280   & 1375$\pm$5     & 1633 $\pm$ 6\\
\midrule
\multirow{3}{*}{1800} & 18    & 7835$\pm$42    & 9717 $\pm$ 50\\
                      & 180   & 859$\pm$4     & 1052 $\pm$ 5\\
                      & 360   & 452$\pm$2     & 550  $\pm$ 3\\
\midrule
\multirow{3}{*}{2000} & 20    & 5375$\pm$29   & 5748 $\pm$ 30\\
                      & 200   & 593$\pm$3     & 635  $\pm$ 3\\
                      & 400   & 315$\pm$2     & 340  $\pm$ 2\\
\midrule
\multirow{3}{*}{2400} & 24    & 2005$\pm$11   & 2146 $\pm$ 13\\
                      & 240   & 235$\pm$1     & 253  $\pm$ 1\\
                      & 480   & 131,7$\pm$0,8   & 142,0$\pm$ 0,8\\
\midrule
\multirow{3}{*}{2800} & 28    & 791$\pm$4     & 850  $\pm$ 5\\
                      & 280   & 102,4$\pm$0,7   & 110,4$\pm$ 0,7\\
                      & 560   & 59,9$\pm$0,4   & 65,1 $\pm$ 0,4\\
\midrule
\multirow{3}{*}{3200} & 32    & 331$\pm$2     & 356  $\pm$ 2\\
                      & 320   & 47,9$\pm$0,4   & 51,9 $\pm$ 0,4\\
                      & 640   & 30,0$\pm$0,1   & 32,6 $\pm$ 0,1\\
\midrule
\multirow{3}{*}{3600} & 36    & 145$\pm$1     & 156  $\pm$ 1\\
                      & 360   & 23,9$\pm$0,1   & 25,8 $\pm$ 0,1\\
                      & 720   & 16,20$\pm$0,08  & 17,51$\pm$ 0,09\\
\midrule
\multirow{3}{*}{4000} & 40    & 67,3$\pm$0,6   & 71,8 $\pm$ 0,7\\
                      & 400   & 13,10$\pm$0,06  & 14,12$\pm$ 0,07\\
                      & 800   & 9,43$\pm$0,04  & 10,25$\pm$ 0,04\\
\midrule
\multirow{3}{*}{4400} & 44    & 34,1$\pm$0,1   & 35,7 $\pm$ 0,1\\
                      & 440   & 7,61$\pm$0,04  & 8,20 $\pm$ 0,04\\
                      & 880   & 5,76$\pm$0,02  & 6,24 $\pm$ 0,03\\
\midrule
\multirow{3}{*}{4800} & 48    & 18,18$\pm$0,03  & 18,91$\pm$ 0,03\\
                      & 480   & 4,70$\pm$0,03  & 5,07 $\pm$ 0,03\\
                      & 960   & 3,77$\pm$0,02  & 4,09 $\pm$ 0,02\\
\midrule
\multirow{3}{*}{5200} & 52    & 10,17$\pm$0,03  & 10,67$\pm$ 0,02\\
                      & 520   & 3,06$\pm$0,02  & 3,34 $\pm$ 0,02\\
                      & 1040  & 2,54$\pm$0,01  & 2,77 $\pm$ 0,02\\
\midrule
\multirow{3}{*}{5600} & 56    & 5,91$\pm$0,01  & 6,40 $\pm$ 0,01\\
                      & 560   & 2,07$\pm$0,01  & 2,28 $\pm$ 0,01\\
                      & 1120  & 1,798$\pm$0,007 & 1,958$\pm$ 0,007\\
\midrule
\multirow{3}{*}{6000} & 60    & 3,528$\pm$0,005 & 4,046$\pm$ 0,007\\
                      & 600   & 1,49$\pm$0,01  & 1,63 $\pm$ 0,01\\
                      & 1200  & 1,304$\pm$0,008 & 1,428$\pm$ 0,005\\
\bottomrule
\end{tabular}}
\caption{\label{tab:wpr_xs} Cross section values and their relative uncertainties obtained with the \textsc{MadGraph5\_MC@NLO} generator for narrow (1\%) and wide (10\%, 20\%, and 30\%) widths $\text{W}^\prime$ boson for different mass hypotheses for both 13 and 14 TeV.}
\end{table}

\section{Method and results}
\label{sec:results}

\begin{figure}[ht]
\centering
\includegraphics[width=70mm]{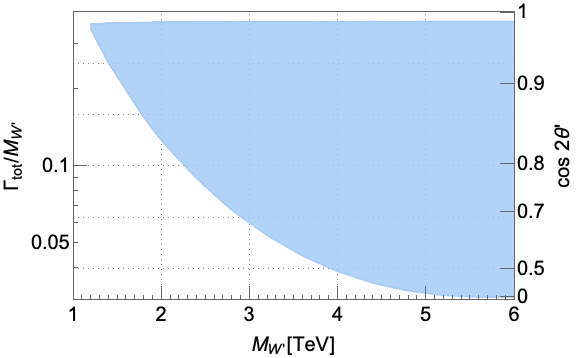}
\caption{Allowed ranges for $\Gamma_{\text{Tot}}/M_{\text{W}^\prime}$ and $\cos\, 2\theta^\prime$ for a fixed $M_{\text{W}^\prime}$ value.}
\label{massa e angolo}
\end{figure}

As discussed in the previous sections, the TF model considered in this work has five free parameters $u$, $v$, $g_{12}$, $g_3$ e $g_0$. The constraints in Eqs.~(\ref{eqg}) and~(\ref{eqn:mass}) allow to reduce the number of free parameters to two: $u$ and $\theta^\prime$.
We require the correction terms due to the TopFlavor model in Eq.\,(\ref{eqn:mass}) for $M_W$ and $M_Z$ to be within the experimental error, thus constraining the allowed values of $\theta^\prime$ as a function of $u$.
We impose a further constraint on the model, by considering that the interactions with the gauge bosons of both $SU(2)$ groups can be perturbatively treated. To accomplish this, we require that $g_{12}^2,g_3^2<4\pi$, obtaining:
\begin{equation}
    0.18<\,\tan \theta^\prime \, < 5.5.
    \label{Constraint1}
\end{equation}
In our analysis, we scan $10^6$ points in the parameter space $(u,\theta^\prime)$, with $\theta^\prime$ satisfying Eq.~(\ref{Constraint1}) and $u > 800\, \text{GeV}$. These conditions ensure that the mass of $\text{W}^\prime$ is larger than $1\,\text{TeV}$. 
For each point of the parameter space we obtain the observable $M_{\text{W}^\prime}$, $\Gamma_{\text{Tot}}$, $\text{Br}(\text{W}^\prime \to tb)$, $\text{Br}(\text{W}^\prime \to \tau \, \nu_\tau)$ from Eqs\,(\ref{eqn:mass}, \ref{gammatot}, \ref{tau}).
We checked the constraints presented in Refs.\,\cite{malkawi1996model, malkawi1999new, Lee:2010zzq} and they do not appear to modify the results mentioned above in the parameter space we considered.
In Fig.\,\ref{massa e angolo} we show all the possible
$\Gamma_{\text{Tot}}/M_{\text{W}^\prime}$ as a function of $M_{\text{W}^\prime}$. The value of $\Gamma_{\text{Tot}}/M_{\text{W}^\prime}$ depends only on $\theta'$, which is the reason why we also show the right vertical axis in terms of $\cos2\theta'$.
The total width can be up to $40\%$ of the $\text{W}^\prime$ boson mass for $\sin \theta'=0.18$.
On the other hand, the minimum value of  $\Gamma_{\text{Tot}}/M_{\text{W}^\prime}$ is obtained for $\theta^\prime=\pi/4$, since for this angle we recover the prediction of the SSM $h_{I,II}^\prime=h_{III}^\prime=g_{SM}$. We note that the total $\text{W}^\prime$ boson width, $\Gamma_{\text{Tot}}$, is approximately proportional to the $\text{W}^\prime$ boson mass.

\begin{figure*}[ht]

\includegraphics[width=70mm]{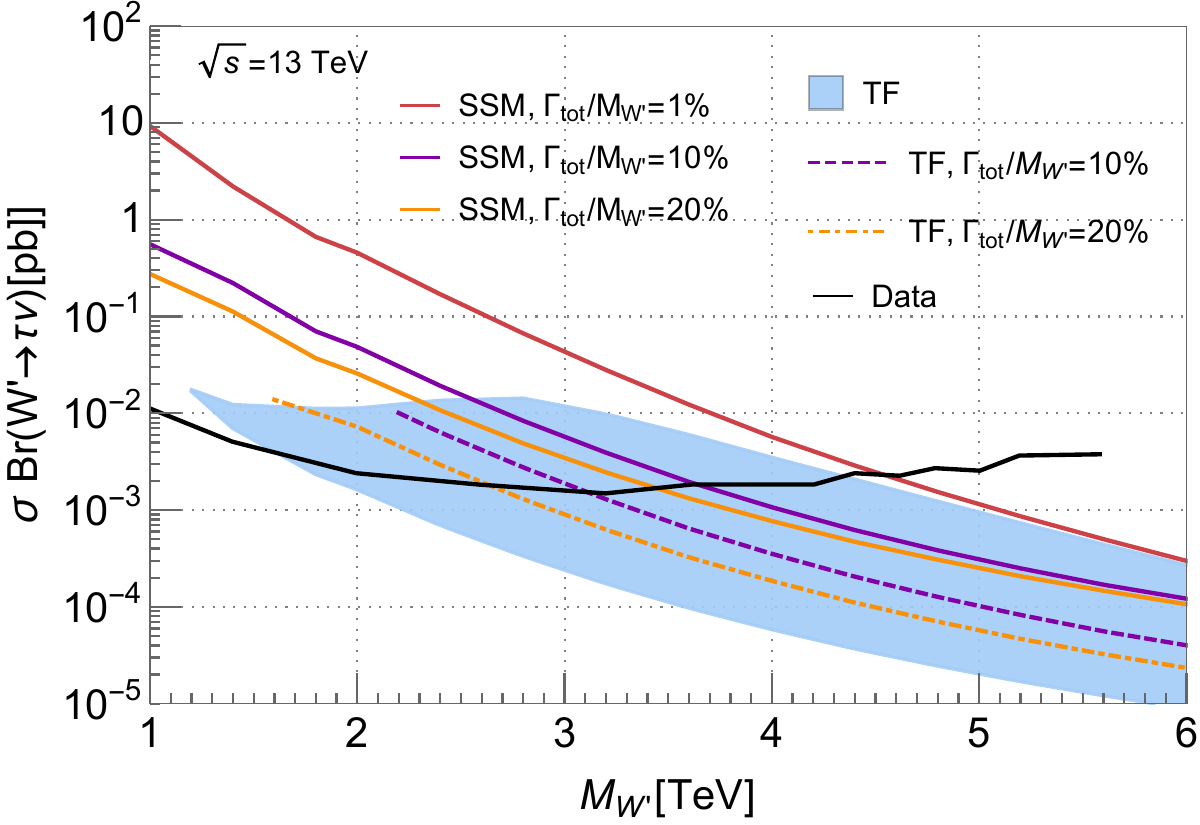}\,\,\,\,\,\,\,\,
\includegraphics[width=70mm]{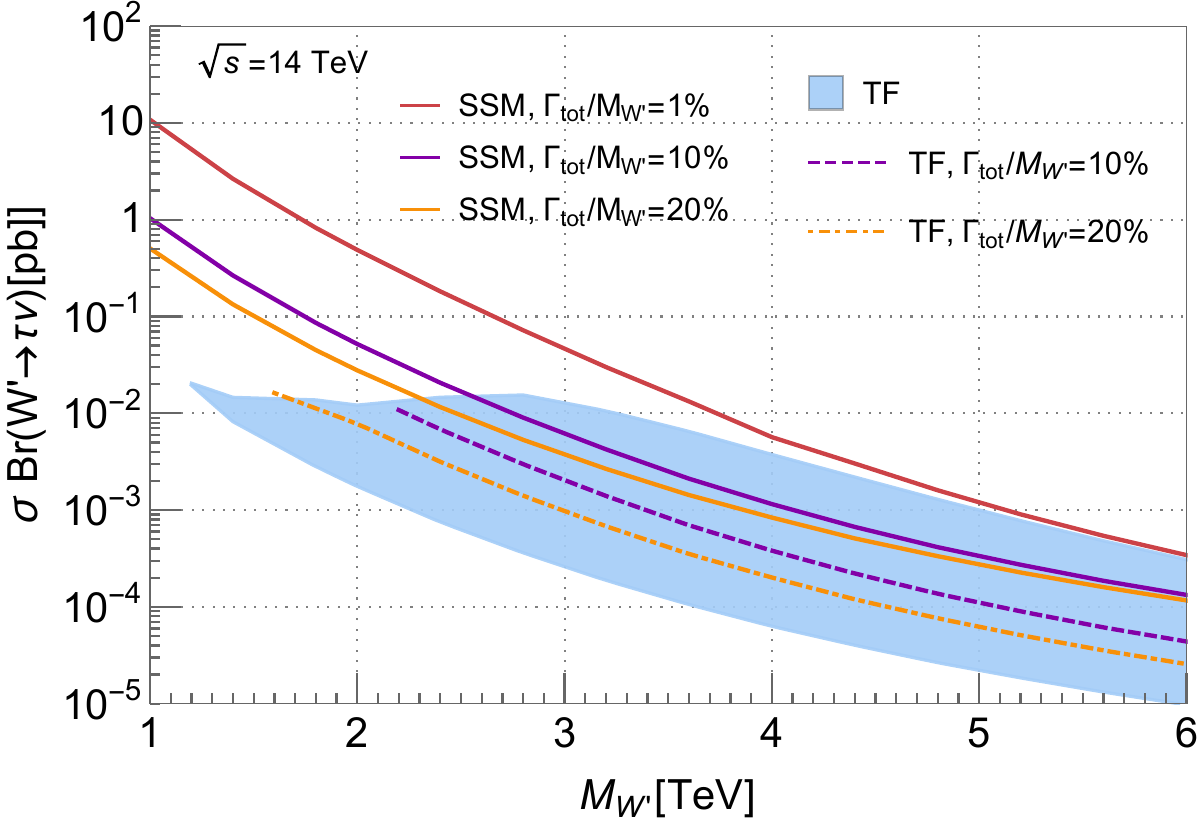}\\
\includegraphics[width=70mm]{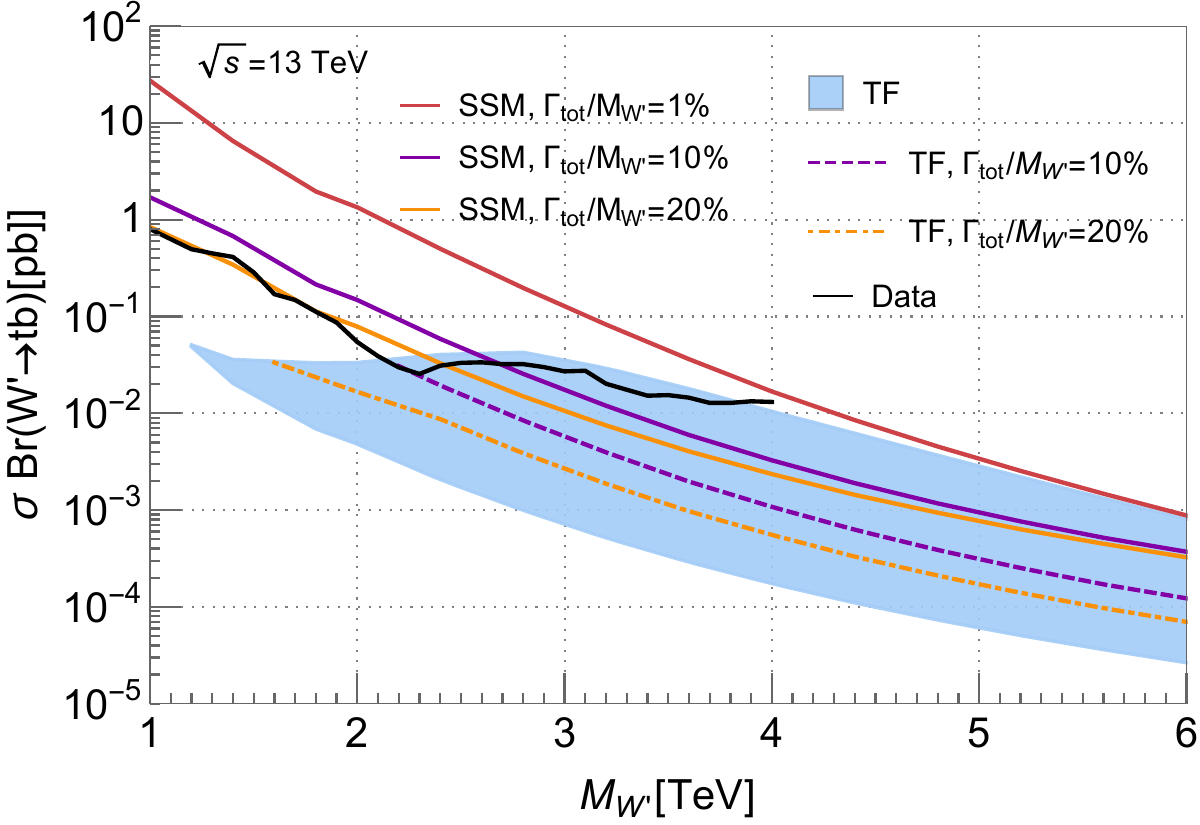}\,\,\,\,\,\,\,\,
\includegraphics[width=70mm]{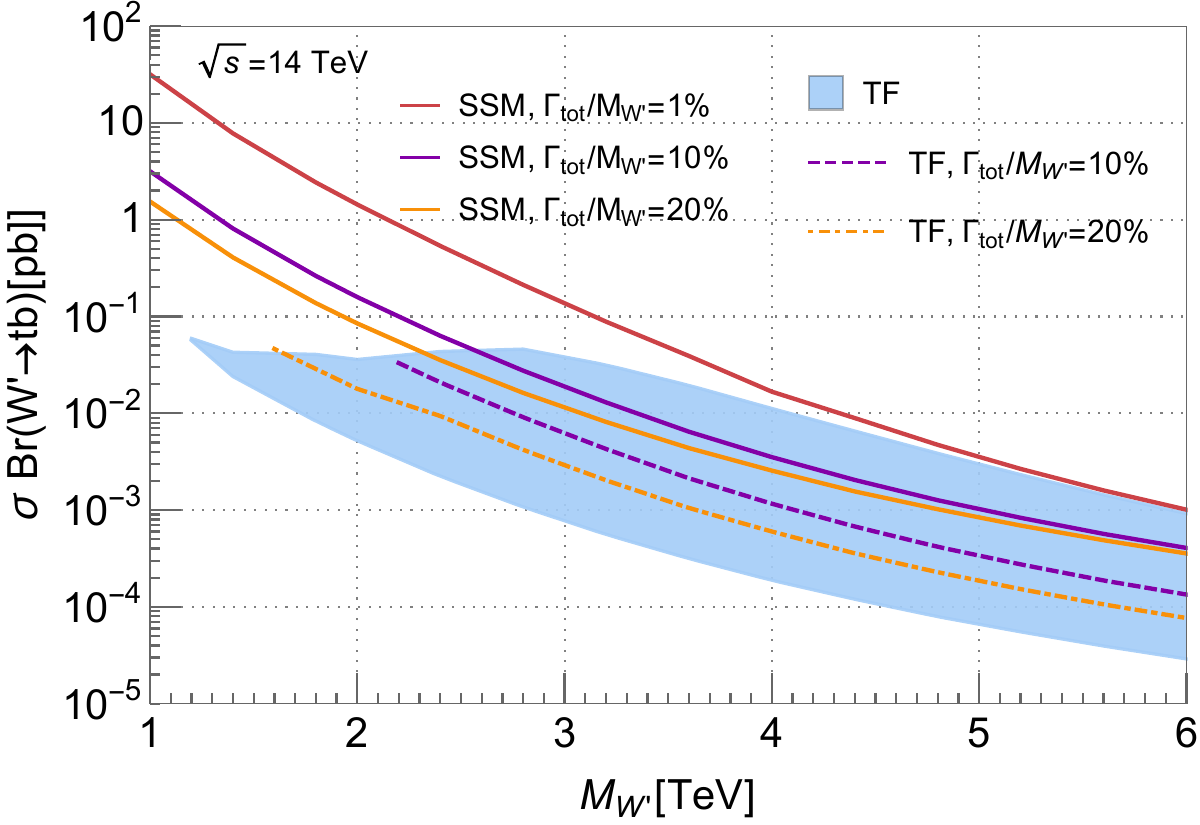}
\caption{Theoretical predictions for $\sigma(pp\to \text{W}^\prime)\text{Br}(\text{W}^\prime \to tb)$ (top panels) and $\sigma(pp\to \text{W}^\prime)\text{Br}(\text{W}^\prime \to \tau\nu_\tau)$ (bottom panels) as a function of $M_{\text{W}^\prime}$ at $\sqrt{s}=13\, \text{TeV}$ (left panels) and $\sqrt{s}=14\, \text{TeV}$ (right panels). The cyan band represents the allowed phase space from the TopFlavor prediction. The dashed violet and orange lines correspond to $\Gamma_{\text{Tot}}/M_{\text{W}^\prime}=0.1,\,0.2$ in the TF model respectively. The red, violet and orange continuous curves are the SSM predictions with $\alpha_L=1$, $\alpha_R=0$ for $\Gamma_{\text{Tot}}/M_{\text{W}^\prime}$ equal to 0.01,0.1, and 0.2 respectively. The black curve was built using the data in Ref.~\,\cite{Sirunyan:2017vkm, Sirunyan_2019}. }
\label{13 TeV}
\end{figure*}
In Fig.\,\ref{13 TeV} we show the cross section times the branching fractions to the third family fermions: 
\begin{equation}
    \begin{split}
        &\sigma(pp\to \text{W}^\prime)\text{Br}(\text{W}^\prime\to tb),\\
        &\sigma(pp\to \text{W}^\prime)\text{Br}(\text{W}^\prime\to \tau \nu_\tau),
    \end{split}
    \label{eqobs}
\end{equation}
as a function of $M_{\text{W}^\prime}$ for a centre-of-mass energy of $\sqrt{s}=13$ and 14 TeV. 
The cyan band represents the parameter space allowed by the TF model. The red line stands for the SSM predicted cross section for $\Gamma_{\text{Tot}}=0.01\,M_{\text{W}^\prime}$. 
For comparison purposes, the predicted cross section for  values of $\Gamma_{\text{Tot}}$ equal to $0.1\,M_{\text{W}^\prime}$, and $0.2\,M_{\text{W}^\prime}$ in both the SSM and TF assumptions are also shown. From these plots is possible to notice that the TF model is up to one order of magnitude smaller than the SSM.

The black lines represent the most recent exclusion limits obtained by the CMS Collaboration~\cite{Sirunyan:2017vkm, Sirunyan_2019} in the context of the W$^\prime$ boson searches in those two decay channels. 
Those plots showcase the  portion of the phase space allowed in the TF model is still not excluded by direct searches. The top-left panel of Fig.~\ref{13 TeV} in particular shows that only values of the $\text{W}^\prime$ mass below 1.6 TeV are excluded in the $\tau\nu$ channels for any value of $\theta^\prime$. Larger values of the mass are possible, with a production cross section times branching fraction of order of magnitude 1 fb, i.e. below the data upper limit. For the $tb$ decay channel, the bottom-left panel shows that a large portion of the phase space is still available, as values of $M_{\text{W}^\prime}$ between 2.2 and 4 TeV are excluded only on for cross sections times branching ratio above order of 10 fb. 
It is also noteworthy that in the allowed region corresponds the total decay width $\Gamma_{\text{Tot}}$ can assume values up to 36\% of the particle mass, which can result in differences in the observable quantities used in physics analyses.

The increase of the centre-of-mass energy at the LHC collider to 14 TeV would increase the production cross sections by a factor $1.1$.

Before conclusions we would like to comment on the recent measurement of the muon magnetic anomaly $g-2$~\cite{Abi:2021gix}, exhibiting a discrepancy with the expected theoretical value at a significance level of $4.2$ standard  deviations\footnote{It is important to stress out that this deviation is not present if one consider the lattice QCD computation, as shown for instance in Ref.\,\cite{borsanyi2020leading}}. From Ref.~\cite{Biggio_2016} one can infer that new physics contributions to $\Delta a_\mu$ in the TF scheme are suppressed for masses of the order of TeV and above.

\section{Conclusions}
\label{sec:conclusions}

LHC searches for new physics also by looking for new charged $\text{W}^\prime$ gauge bosons through a variety of final states including leptons~\cite{Sirunyan:2018mpc,Sirunyan:2019vgj,Sirunyan_2019,Sirunyan:2017vkm} or quarks~\cite{Atlas_wtolep,Aaboud:2017yvp,Aaboud:2018vgh,Aaboud_2019,Aaboud:2018juj,CMS:2021mux}. 
Models based on the Sequential Standard Model are often used as benchmarks for such analyses that look after final states where the $\text{W}^\prime$ decays to fermions.
Even if the SSM incorporates a wide variety of models, other theories predicting new heavy bosons might have realizations that are not allowed in the SSM, and that have not been yet excluded by data. We provide the  particular case of the TopFlavor model, where a phenomenological study is performed, and the allowed parameter range is explored and expressed in terms of observable quantities at the LHC, like the new particle mass or width. For fixed values of $\Gamma_\mathrm{tot}/M_{\text{W}^\prime}$, the study resulted in:
\begin{equation*}
    10^{-1}\lesssim\frac{\left[\sigma(pp\to \text{W}^\prime)\cdot \text{Br}(\text{W}^\prime\to ff)\right]_{\rm \text{TF}}}{\left[\sigma(pp\to \text{W}^\prime )\cdot \text{Br}(\text{W}^\prime\to ff)\right]_{\rm \text{SSM}} }\lesssim 1\,.
\end{equation*}

This shows that a more systematic study of $\text{W}^\prime$-models for LHC search is required.

While standard CMS and ATLAS analyses have already performed thorough searches for a $\text{W}^\prime$ in the $tb$ and in $\tau \nu$ final states,  in the TopFlavor model such measurements provide lower limits for the $M_{\text{W}^\prime}$ mass of the order of $2\,\text{TeV}$. A significant portion of the phase space is therefore still allowed in the TopFlavor model.

\section*{Acknowledgments}
Work supported by the Italian grant 2017W4HA7S “NAT-NET: Neutrino and Astroparticle Theory Network” (PRIN 2017) funded by the Italian Ministero dell’Istruzione, dell’Universit\`{a} e della Ricerca (MIUR), and Iniziativa Specifica TAsP of INFN.

\bibliographystyle{unsrt}
\bibliography{databaseWprime} 
\end{document}